**Title:** EFFICIENT ASSEMBLY OF BOLTED JOINTS UNDER EXTERNAL LOADS USING NUMERICAL FEM

**Authors:** Ibai Coria[1], Iñigo Martín[1], Abdel-Hakim Bouzid[2], Iker Heras[1], Mikel Abasolo[1]

1. Department of Mechanical Engineering, ETSI Bilbao, University of the Basque Country (UPV/EHU), Alameda Urquijo s/n, 48013, Bilbao, Spain.
2. Department of Mechanical Engineering, École de technologie supérieure, University of Quebec, 1100, rue Notre-Dame Ouest, H3C 1K3, Montreal, Canada.

[*] Corresponding author:
Ibai Coria
E-mail address: ibai.coria@ehu.eus
Tel.: +0034 946 01 7433.

## Abstract

Several factors can cause leakages in a bolted flange connection, external loads being the most important one. External loads such as those produced by misalignment introduce an external axial load combined with a bending moment which could lead to either an excessive compression of the gasket and its subsequent crushing, and/or a separation of the flanges causing leakage failure. Several studies have been carried out to study these prejudicial effects, but no practical solutions are proposed to compensate for them. In this sense, this work presents a numerical methodology that iteratively calculates the non-uniform bolt tightening load distribution to achieve a uniform gasket stress distribution that improves the leakage performance of bolted joints subjected to external loads.

## 1. Introduction

In the Oil & Gas installations, bolted flange connections are widely used to connect pressure vessels and pipes together. This is because they are practical to use during the assembly and disassembly process and, consequently they are very attractive for maintenance operations [1]. These connections are either standards according to ASME B16.5 or API or specifically designed according to appendix 2 of the ASME code section VIII division 2 [2].

Several research works have proposed methods for the optimization of the assembly process in order to obtain more efficient tightening sequences [3-6]. The objective of these methodologies is to achieve a uniform contact pressure distribution around the gasket periphery by achieving a uniform bolt load distribution at the end of the tightening process during assembly. However, these methods assume ideal conditions in the installations as no external loads produced by misalignment or during operation are taken into account. In order to overcome the non-uniform gasket contact distribution produced by external bending loads acting on the bolted joint during operation, one of the solutions would be to tighten the bolts non-uniformly. The amount of initial tightening of each bolt is determined such that the gasket stress distribution would be uniform.

One of the most important sources of external loads is flange misalignment, caused by tolerances, fabrication, assembly, and installation practices [7]. Misalignment introduces significant external axial and bending loads in the bolted joint; radial loads are compensated by the frictional loads between the bolted joint members [8]. An external axial load can cause a gasket stress decrease while an external bending moment causes one side of the gasket to increase and the other side to decrease. As a result leakage failure occurs due to crushing or not enough gasket stress to seal the bolted joint leading to shut down to take corrective actions, with the subsequent delay in production and economic losses [9].

Few research studies are focused on the influence of external loads on the gasket contact stress quantifying theoretically and/or experimentally the resulting amount of leakage. Bouzid [10] developed an analytical model to obtain the leak rate under an external bending moment. Koves suggested a code like formulae to account for external bending load in flange design [9]. Abid et al. [11] determined experimentally the load capacity of the joint for safe performance when external axial and bending moment are applied. These studies determine the maximum axial/bending loads that a bolted joint with uniform load distribution can withstand without significant leakage. None of the works found in the literature covers the initial bolt-up compensation of external bending loads to achieve uniform gasket contact stress during operation [8,12].

To fulfill this need, the present work focuses on the development of a new methodology in order to avoid failures due to external loads. As mentioned before, in ideal conditions with no external loads a uniform bolt load distribution provides a uniform gasket stress distribution. On the contrary, under external loads, a non-uniform bolt load distribution may be useful in order to achieve a uniform gasket stress distribution after assembly. Knowing the amount of external loads, the proposed methodology is used to determine the specific non-uniform bolt load distribution by means of an iterative algorithm programmed in conjunction with a superelement-based Finite Element (FE) model. The developed model is validated experimentally using data from previous work [10]. Additionally, this strategy allows increasing the admissible maximum axial/bending loads predicted by previous works [10-11]. The approach can be exploited by assemblers to ensure safe performance of bolted joints subjected to external loads.

## 2. FEM based methodology

Flange misalignments generate a bending moment and an axial load in a bolted flange joint. For certain misalignment deviations, the magnitude of the resulting loads depends on the stiffness of the installation [13]. Like external loads generated by dead weight, temperature expansion and others loads produced by misalignments must be estimated or measured in field. Once these loads are known, they are introduced in the FE model of the bolted joint to estimate the non-uniform initial bolt-up loading. Using an algorithm programmed via APDL scripts, the methodology performs a series of iterative analyses in order to calculate the non-uniform bolt load distribution that would achieve a uniform gasket contact stress distribution after assembly and/or during service.

In order to explain in detail this new methodology, the superelement-based model and its experimental validation are laid out before presenting the algorithm of the methodology.

### 2.1  Superelement-based FE model

Because of the iterative nature of the methodology, a powerful FE tool was required in order to perform a large number of analyses in an efficient way. These advantages are provided by the superelement technique, which reduces the computational cost with no loss of accuracy [14].

The parametric FE superelement-based model was developed in Ansys® Mechanical APDL. For that purpose, first a conventional FE model of a NPS 4 class 150 flange with a compressed fiber gasket was generated: only one half of the joint (i.e. one flange and half gasket) was modeled due to the symmetry of the system [4,15-16], and the bolt bodies were modeled by beam elements (with the same stiffness as the bolts) attached to the flange via rigid beams simulating bolts head [17-19]; it was verified that beam modeling and conventional 3D bolt modeling provided almost identical results, being the first alternative more efficient. Finally, a remote node was defined in order to apply the external loads, which was connected to the upper face of the pipe by means of rigid beams. The model, shown in Figure 1, was meshed mainly with high-order solid elements with a total of 322,899 DoF.

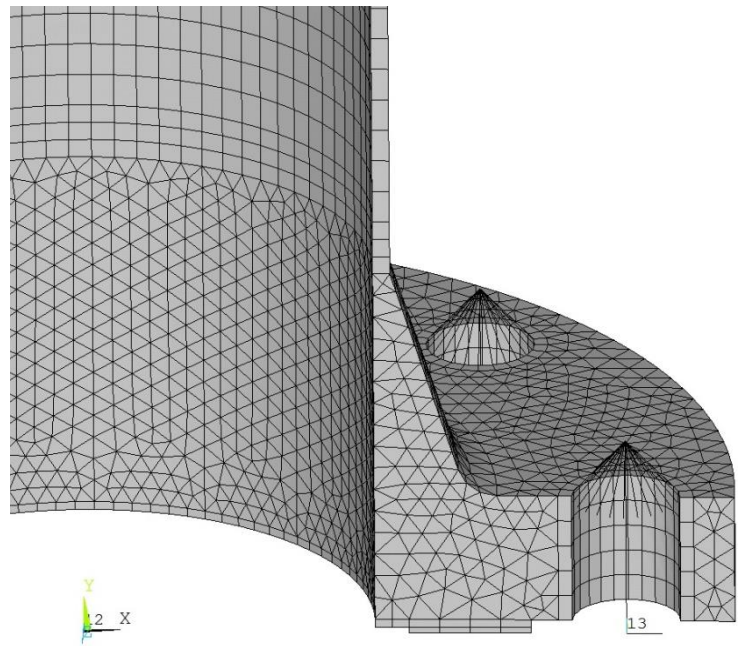

Figure 1. Conventional FE model

As a second step, the superelement-based FE model was created from the conventional model. The superelement consists of the flange, the beams that connect the pipe to the load application node, and finally the bolt head beams, so the gasket and the bolt body beams were not included; these latter were excluded because no nonlinearities sources (gasket material) or load application points (pretension sections) are allowed inside the superelement, respectively. The master nodes of the superelement, that is the nodes that connect the superelement with the rest of the model or the nodes where loads or boundary conditions are applied, are the nodes that connect the bolt bodies to their corresponding heads, the nodes that connect the flange to the gasket, the load application node and the nodes with a boundary condition. Figure 2 shows the master nodes of the superelement. The boundary conditions of the model are the ones related to the symmetry of the system; the gasket lower nodes are clamped, and only planar movement is allowed to the lower nodes of the beam bodies. As a result, the superelement-based FE model of Figure 3 is obtained, which has 3003 DoF (much less than the 322,899 DoF of the conventional model). Accordingly, the computational cost is significantly lower.

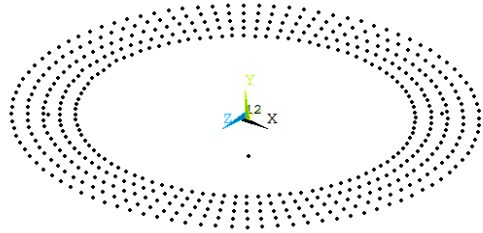

Figure 2. Masternodes of the superelement

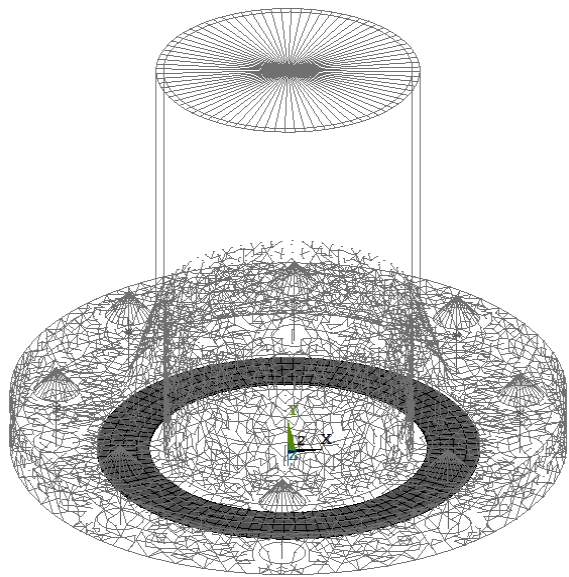

Figure 3. Superelement-based FE model

## 2.2 Experimental Validation of the Superelement-Based FE Model

The superelement-based model was validated by comparing its results with the results obtained by Bouzid in the experimental test bench used in [10] (see Figure 4). In order to replicate the experimental test and compare the results, the same geometry and material properties of the test bench were imposed to the superelement-based model. Therefore, a NPS 4 class 150 flange was modeled with a compressed fiber gasket. Figure 5 shows the compression curve of the gasket, obtained from a compression test in a ROTT machine (Room Temperature Operational Tightness Test) [20].

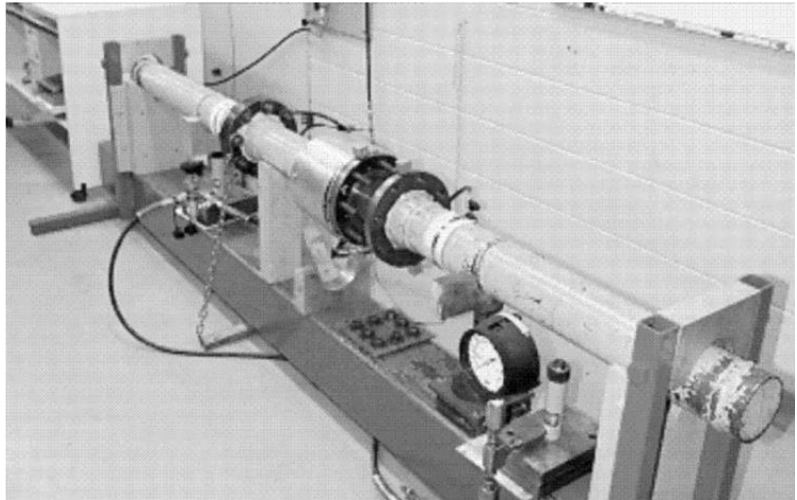

Figure 4. Experimental test bench

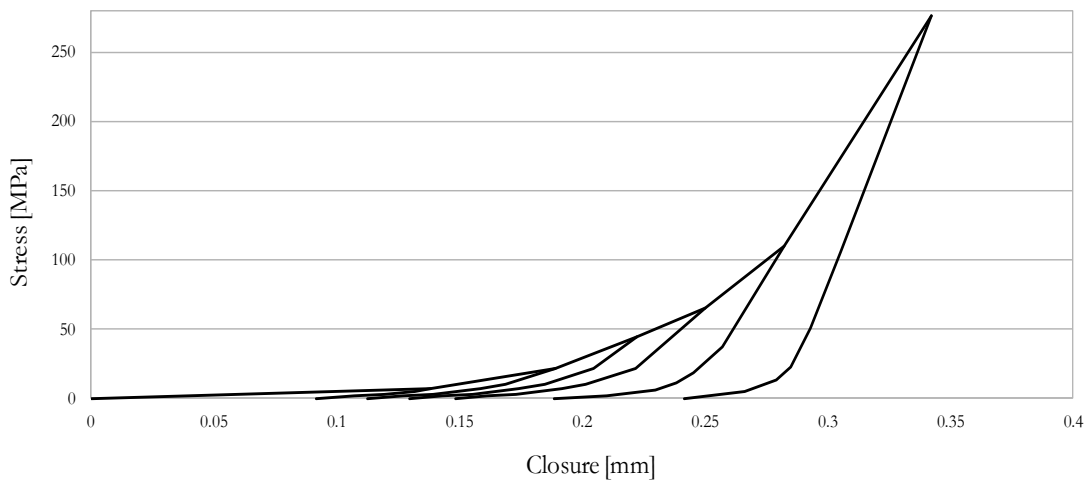

Figure 5. Compression curve of the gasket

Test bench was only able to apply bending moments in the joint, so the validation was only performed for this load case. Nevertheless, accuracy level for axial load case should be similar. Among the different experimental tests in [10], most critical one was chosen to validate the superelement-based model, which sets a bolt-up stress of 175 MPa in the first load step and a 9600 N·m bending moment in the second load step. Figure 6 compares bolt load variations due to the bending moment in both models; it can be appreciated that the results are in good agreement. Therefore, it can be stated that the superelement-based FE model developed in this work provides accurate results with a very low cost compared to conventional FE models and experimental tests.

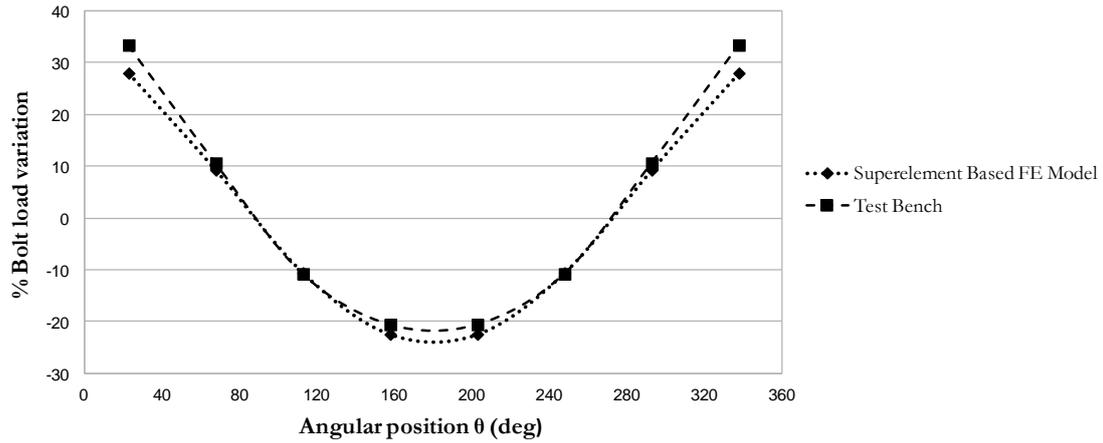

Figure 6. Bolt preload variations in the superelement-based model and in the experimental test.

## 2.3 Algorithm

The basis of the methodology is the iterative algorithm programmed via APDL scripts in the superelment-based FE model, whose background will be explained in this section. As mentioned, the purpose of the algorithm is to achieve the non-uniform bolt load distribution that provides an optimum gasket stress distribution as similar as possible to the ideal (no external loads) case, since exactly the same distribution is unachievable. Thus, the gasket stress distribution will be considered optimal when the gasket stress value in the external radius takes the same value as in the ideal case. This was considered to be an equivalent situation in terms of sealing capability, however different assumptions could be adopted in the algorithm depending on the criterion of the assembler.

Figure 7 shows an illustrative flowchart with the steps of the algorithm. The first step consists on carrying out an analysis that corresponds to the ideal assembly conditions, i.e. with no external loads; in this case, a uniform gasket stress distribution is achieved by applying a uniform bolt load distribution. The gasket stress value in the external radius obtained in this step will be considered as the target stress.

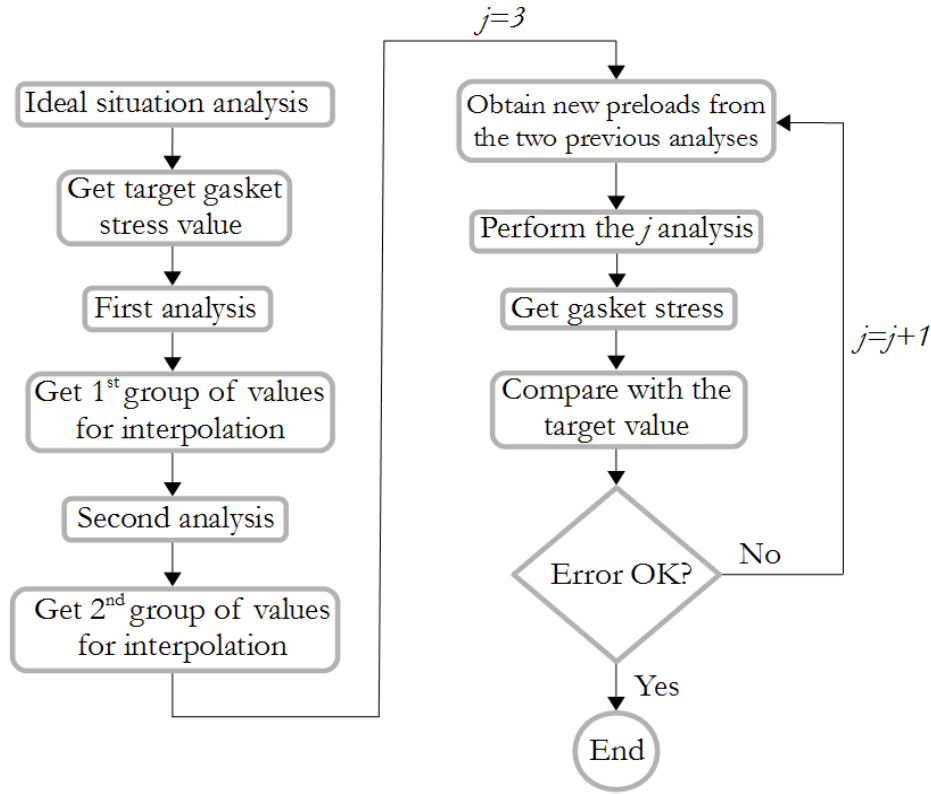

Figure 7. Flowchart of the algorithm

As it will be explained later, the algorithm works with linear interpolations during the iterative process, so two initial sets of preloads and gasket contact pressures are needed as a starting point. Obviously, the convergence of the algorithm is dependent of the correct election of these two analyses.

In the first analysis of the iterative process (see Figure 7), the ideal situation analysis is used but increasing 50% the bolt preloads and applying the external loads. The bolt load increase reduces the possibility of a contact loss in the gasket (contact loss introduces a high nonlinearity in the iterative process, making it more difficult to converge). The gasket stress and bolt preload values obtained in this analysis are used as the 1$^{st}$ group of interpolation. In the second analysis, bolt preloads are modified so as to compensate for the external misalignment loads; note that the ideal situation is not recovered, however it is a good approach. Equation (1) determines bolt preload variation in each bolt to achieve this situation, according to the schematic representation shown in Figure 8:

$$\Delta P_i = \frac{F_A}{n} + \frac{2 \cdot F_m \cdot M \cdot \cos\theta_i}{n \cdot R} \tag{1}$$

Where subscript $i$ denotes the bolt number, $n$ is the number of bolts, $M$ is the bending moment, $F_A$ is the axial load, and $R$ and $\theta$ are the radius and the angular position of the bolt. $F_m$ is a correction rigidity factor, which is explained in [9-10]. From this analysis the 2$^{nd}$ group of interpolation is obtained.

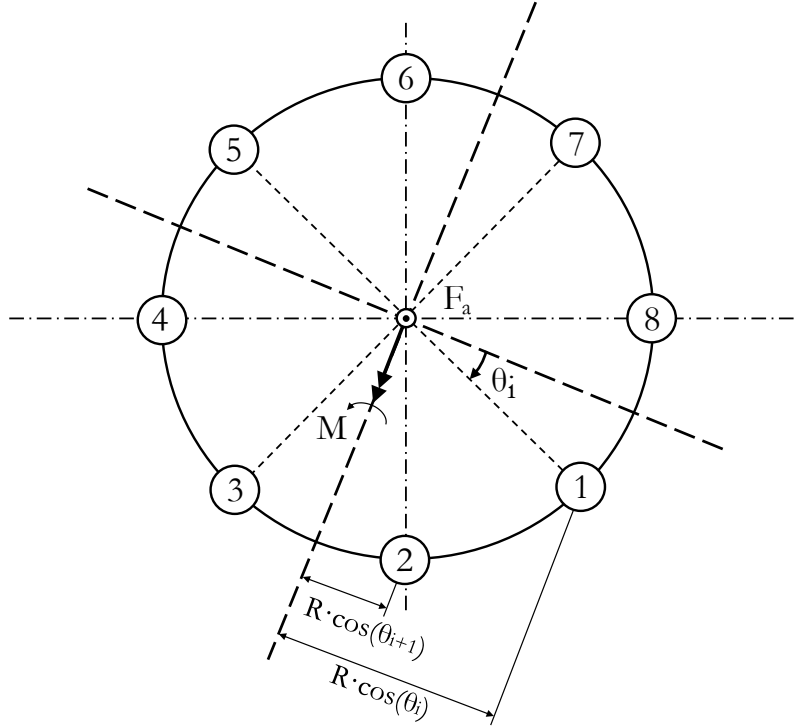

Figure 8. Parameters of the equation (1)

Once the two initial approximations are performed, the iterative process of Figure 7 starts. In this process, each circumferential sector of the joint is studied separately, assuming that the gasket contact pressure in each sector is only influenced by the preload of the corresponding bolt; even though this is not realistic due to elastic interaction [21-24], it simplifies the iterative process (in exchange more iterations are needed but the same final results are obtained). Thus, performing a linear interpolation for every sector, a new set of bolt preloads is obtained, which will be closer to the final solution:

$$P_i^j = P_i^{j-1} - \frac{P_i^{j-1} - P_i^{j-2}}{Gs_i^{j-1} - Gs_i^{j-2}} \cdot (Gs_i^{j-1} - Gs_{target}) \qquad (2)$$

Where superscript $j$ denotes the iteration number, $Gs$ is the gasket stress, $P$ is the bolt preload, and finally $Gs_{target}$ is the target gasket stress value. The iterative process will be repeated until the error is within acceptable limits.

## 3. Results and discussion

The joint of the experimental test bench (see Figure 4) was used to validate the methodology. Apart from bolt preload and bending moment, an axial misalignment load was also applied, as summarized in Table 1.

| Bolt Preload | 175 [MPa] |
|---|---|
| Bending Moment | 9600 [N·m] |
| Axial load | 200 [kN] |

Table 1. Conditions of the study

According to the flowchart in Figure 7, the first step consists of obtaining the target gasket stress value by simulating an ideal (no external loads) assembly with uniform bolt load distribution. Figure 9 shows the resulting gasket stress distribution where, as previously explained, the stress of the external radius will be used to define the target stress value, in this case 36.4 MPa.

When the bending and axial loads of Table 1 are applied to the situation of Figure 9, the gasket stress distribution of Figure 10 is obtained. It can be observed that gasket stress increases in some zones and decreases in others, with a huge loss of contact. In order to recover a gasket stress distribution as similar as possible to the ideal one (optimum gasket stress distribution), the methodology was applied. For such purpose, the procedure described in Figure 7 was programmed in Ansys via APDL scripts. Figure 11 shows the results for intermediate analyses ($5^{th}$, $10^{th}$ and $15^{th}$ iterations) together with the final stress distribution ($21^{st}$ iteration), where the target stress value of 36.4 MPa (±3%) all along the external radius is achieved. Figure 12 shows the nodal gasket stress in the external radius extracted from Figures 9 (ideal situation), 10 (under external loads) and 11 (applying the methodology). Figure 13 shows the evolution of the gasket stress distribution in the external radius throughout the iterative process (mean, minimum and maximum values) until target gasket stress value is achieved all around the gasket. Table 2 shows obtained bolt preloads with the methodology in order to achieve the final optimum situation so as to compensate for external loads. The whole iterative process lasts only 3 minutes. Finally, as a further validation, bolt loads of Table 2 were applied to the conventional FE model of Figure 1, but modeling the bolts as solid elements instead of the aforementioned beam-based equivalent bolt models. As expected, the same results as in the superelement model were obtained, being the average error only 1.9%.

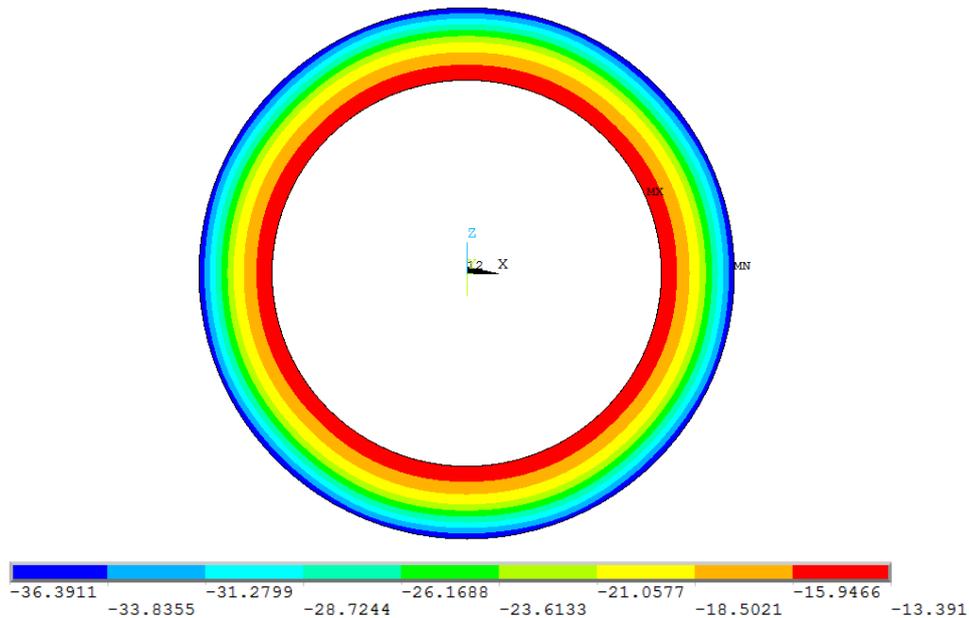

Figure 9. Gasket contact stress (ideal situation)

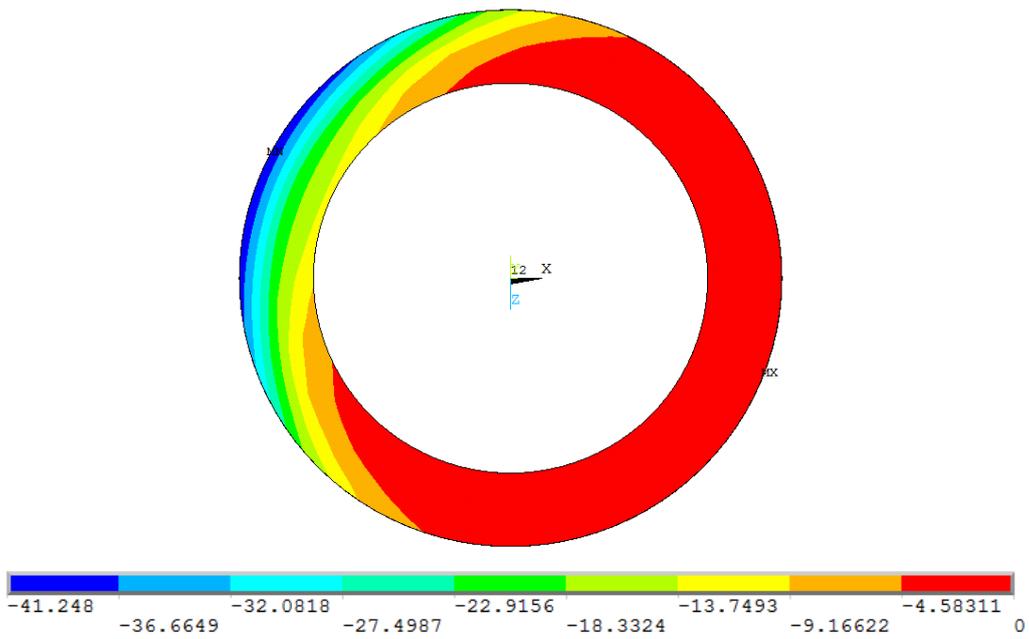

Figure 10. Gasket contact pressure when external loads are applied

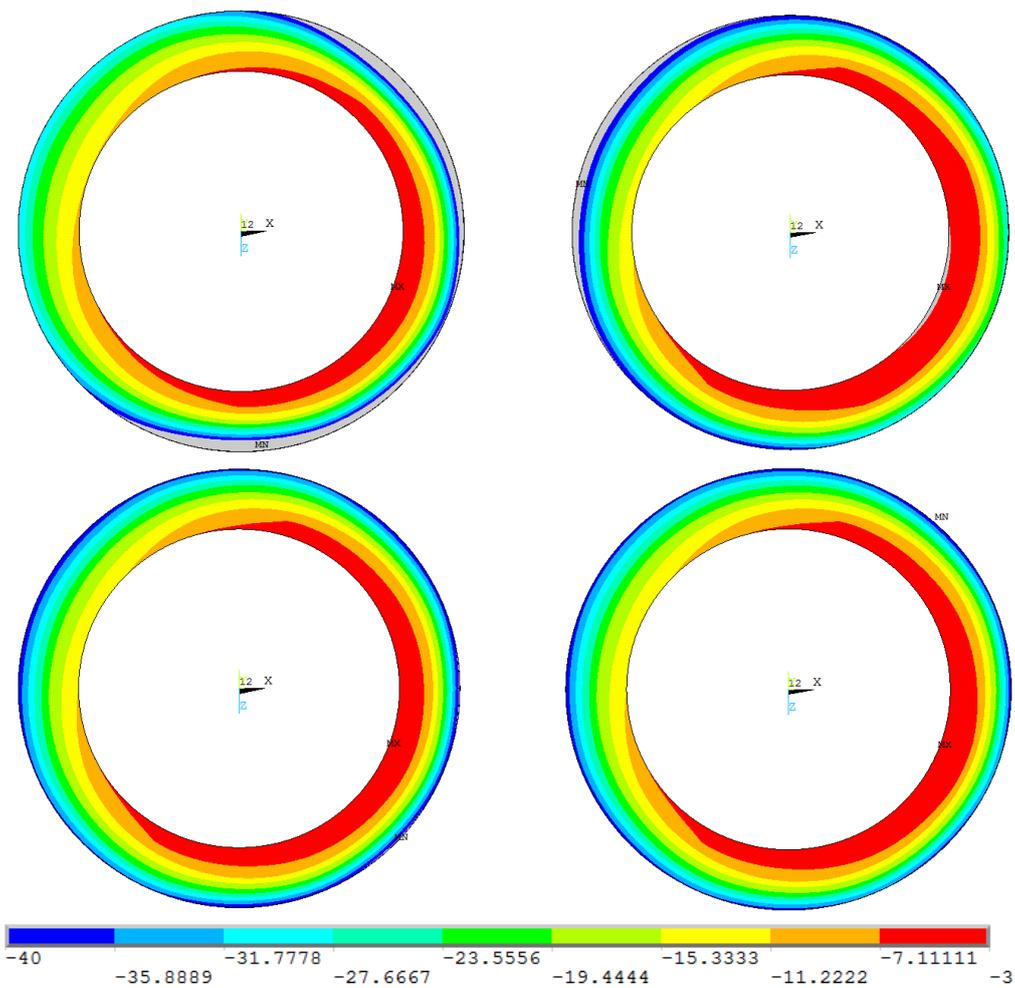

Figure 11. Gasket contact pressure: 5$^{th}$ iteration (top left), 10$^{th}$ iteration (top right), 15$^{th}$ iteration (bottom left), 21$^{st}$ iteration (bottom right). Grey zones are out of scale.

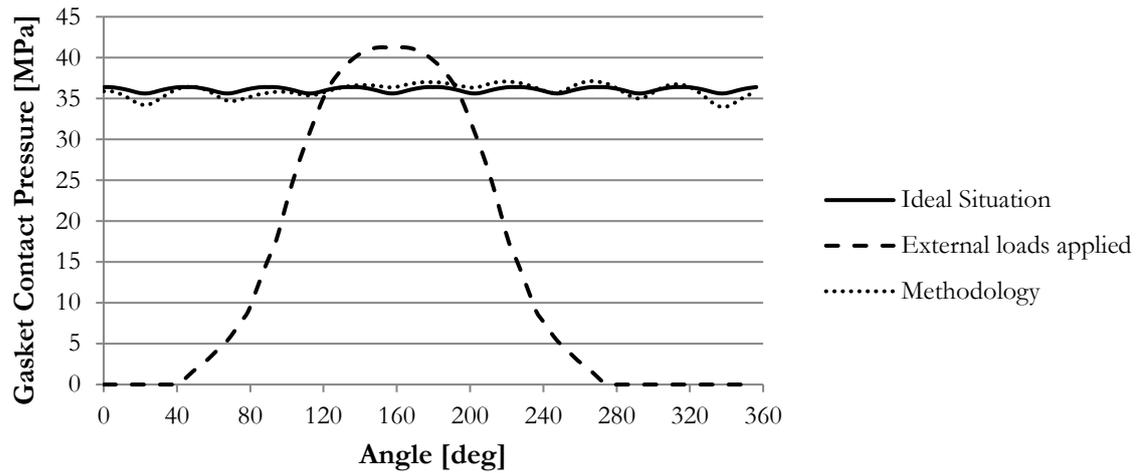

Figure 12. Gasket contact pressure along the gasket external radius in the different situations

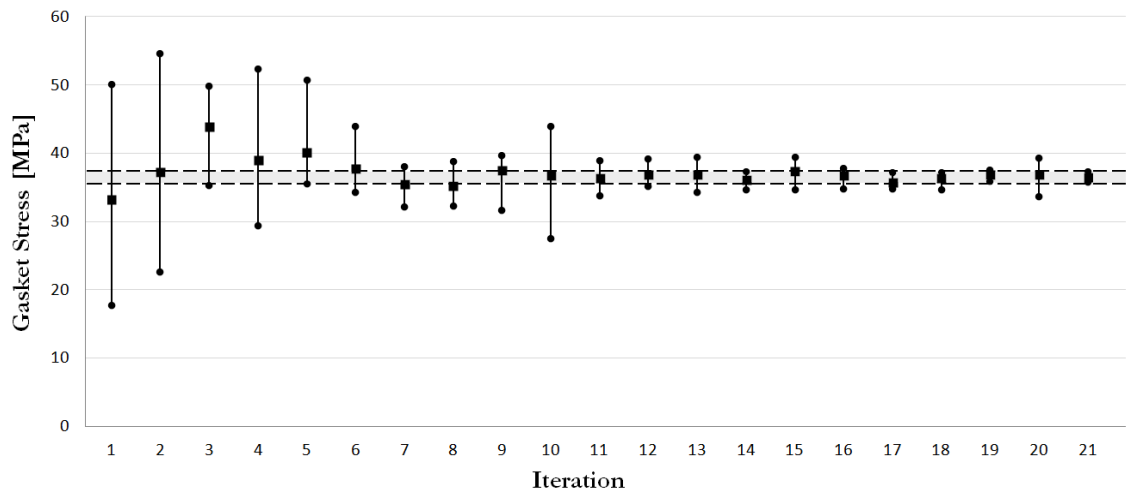

Figure 13. Gasket stress in the external radius along the iterative process: mean, maximum and minimum values. Grey band corresponds to target value 36.4 (±3%) MPa

| Bolt number | 1 (22.5°) | 2 (67.5°) | 3 (112.5°) | 4 (157.5°) | 5 (202.5°) | 6 (247.5°) | 7 (292.5°) | 8 (337.5°) |
|---|---|---|---|---|---|---|---|---|
| Bolt load [MPa] | 432.1 | 351.3 | 239.7 | 165.7 | 165.5 | 233.6 | 352.6 | 425.9 |
| Bolt load variation [%] | 146.9 | 100.7 | 36.9 | -5.3 | -5.3 | 33.4 | 101.5 | 143.4 |

Table 2. Bolt preloads that provide optimum gasket stress distribution under external loads according to the methodology

## 4. Conclusions

Axial and bending loads modify the gasket stress distribution in bolted joints, and as a consequence leakage failure may occur due to crushing or not enough gasket stress to seal the bolted joint. This work presents a methodology that calculates the non-uniform bolt load distribution that is necessary to recover a gasket stress distribution as similar as possible to the ideal situation, where no external loads are applied.

For that purpose, a highly efficient superelement-based FE model was generated and experimentally validated; then, an algorithm was programmed via APDL scripts in order to obtain iteratively the final optimum solution. A particular case was studied, where a gasket stress value within a tolerance of ±3% was obtained in just 3 minutes.

## Acknowledgments

The authors wish to acknowledge the financial support of the Spanish Ministry of Economy and Competitiveness (MINECO) through Grant number DPI2017-85487-R (AEI/FEDER, UE) and the Basque Government through Grant number IT947-16.